# Mesoscale tissue properties and electric fields in brain stimulation: Bridging the macroscopic and microscopic scales using layer-specific cortical conductivity

Boshuo Wang[1], Torge H. Worbs[2,3], Minhaj A. Hussain[4], Aman S. Aberra[5,6], Axel Thielscher[2,3], Warren M. Grill[4,7,8,9], and Angel V. Peterchev[1,4,7,8,*]

[1] Department of Psychiatry and Behavioral Sciences, Duke University, Durham, NC, USA
[2] Section for Magnetic Resonance, DTU Health Tech, Technical University of Denmark, Kgs Lyngby, Denmark
[3] Danish Research Centre for Magnetic Resonance, Department of Radiology and Nuclear Medicine, Copenhagen University Hospital Amager and Hvidovre, Copenhagen, Denmark
[4] Department of Biomedical Engineering, School of Engineering, Duke University, Durham, NC, USA
[5] Department of Biological Sciences, Dartmouth College, Hanover, NH, USA
[6] Department of Molecular and Systems Biology, Dartmouth College, Hanover, NH, USA
[7] Department of Electrical and Computer Engineering, School of Engineering, Duke University, Durham, NC, USA
[8] Department of Neurosurgery, School of Medicine, Duke University, Durham, NC, USA
[9] Department of Neurobiology, School of Medicine, Duke University, Durham, NC, USA
* Corresponding author: angel.peterchev@duke.edu

## Abstract

**Objective.** Accurate simulations of electric fields (E-fields) in neural stimulation depend on tissue conductivity representations that link underlying microscopic tissue structure with macroscopic assumptions. Mesoscale conductivity variations can produce meaningful changes in E-fields and neural activation thresholds but remain largely absent from standard macroscopic models. Conductivity variations within the cortex are expected given the differences in cell density and volume fraction across layers. We sought to estimate layer-specific conductivity within the cortex using computational models.

**Approach.** We review recent efforts modeling microscopic and mesoscopic E-fields and outline approaches that bridge micro- and macroscales to derive consistent mesoscale conductivity distributions. Using simplified microscopic models, effective tissue conductivity was estimated as a function of volume fraction of extracellular space, and the conductivities of different cortical layers were interpolated based on experimental volume fraction.

**Main Results.** The effective tissue conductivities were monotonically decreasing convex functions of the cell volume fraction. They followed the theoretical power formula only for small volume fractions (< 20%) and deviated to higher values for larger volume fractions. With decreasing cell volume fraction, the conductivity of cortical layers increased with depth from layer 2 to 6. Although the variation of conductivity within the cortex was small when compared to the conductivity of extracellular fluid (9% to 15%), the conductivity difference was considerably larger when compared between layers, e.g., with layer 3 and 6 being 20% and 50% more conductive than layer 2, respectively.

**Significance.** The review and analysis provide a foundation for accurate multiscale models of E-fields and neural stimulation. Overcoming technical limitations that microscale and macroscale methods face will allow validation to arrive at consistent conductivity distributions. Using layer-specific conductivity values within the cortex could improve the accuracy of estimations of thresholds and distributions of neural activation in E-field models of brain stimulation.

# 1. Introduction

Accurate modeling of the electric fields (E-fields) generated by brain stimulation is crucial for estimating the stimulation dose [1], understanding the neural response [2], and improving performance via computational optimization [3]. Conventional E-field simulations assume homogeneity of tissue electrical properties on the macroscopic scale that represents an average of the microscopic cellular composition [4]. Unlike previous microscopic E-field simulations restricted to only a few simplified cells [5], the most recent models captured much larger tissue volumes up to one cubic millimeter with morphologically-realistic cells reconstructed from electron microscopy [6]. The resulting models enabled the evaluation of the effects of microscopic E-field perturbations on neural activation thresholds in response to, for example, transcranial magnetic stimulation [6–9] and sparked constructive debates about the uses and limits of microscopic models [4,8–10]. We summarize and extend these discussions and outline a path toward coherent multiscale modeling frameworks of E-fields and neural activation generated by brain stimulation that span the micro-, meso-, and macroscales [4].

We start by defining the range covered by the mesoscale in the context of brain stimulation and describe why conductivity variations on this scale influence stimulation thresholds. We then summarize the status and challenges of macro- and microscale approaches that have potential to inform mesoscopic conductivity estimates. In the main part of the paper, we tackle the current lack of suitable 3D microscopic datasets by introducing an approach to estimate layer-specific mesoscopic conductivities from simulations of 3D blocks of simplified microscopic cellular representations informed by cell volume fraction data. Finally, we conclude by briefly summarizing the key points of this study.

# 2. Theoretical framework and previous work

## *2.1. Mesoscale and its relevance for brain stimulation*

Microscale corresponds to sizes of cellular and sub-cellular structures in the range of nanometers to tens of micrometers. Macroscale, in the broader sense, refers to any tissue volumes larger than microscale for which the underlying cellular composition can be (locally) homogenized. Depending on the model size and stimulation modality, macroscale can be considered on the order of several millimeters in size or larger, e.g., full head model for transcranial electromagnetic stimulation based on millimeter-resolution imaging data. In this context, the two orders-of-magnitude between the microscale and macroscale are considered the mesoscale. In other cases, e.g., deep brain stimulation or vagus nerve stimulation with implanted electrodes, macroscale corresponds to sizes of the electrode and subcortical nuclei or fascicle, which is similar to mesoscale for a full head model, and there may be a less clear distinction between macro- and mesoscales.

Head models implemented with the finite- or boundary-element method typically assign constant conductivity to each tissue volume, constrained by computational cost and the lack of spatially distributed measurements. Macroscopic E-field models almost always use homogenous conductivities despite being capable of simulating inhomogeneity, and the term "macroscopic" may be conflated with "homogenous" in modeling work [9]. Mesoscale conductivity also represents averages of microscopic cellular composition, in contrast to microscopic models that explicitly represent underlying cellular structures. Measurements of mesoscopic inhomogeneities, such as layer- or depth-dependent conductivities (or, equivalently, resistivities) in the cerebral, cerebellar, and hippocampal cortices [11–14] and in the retina [15–17], are available but rarely modeled [18,19]. An important exception is white matter anisotropy modeled from diffusion tensor imaging data: the conductivity tensor varies on the mesoscale, although its magnitude is usually derived from a fixed equivalent isotropic conductivity value [20]. Major challenges also remain regarding partial volume effects at tissue boundaries due to the limited spatial resolution and lack of consensus on diffusion-conductivity mappings [20,21].

Mesoscopic conductivity variations in macroscopic models are expected to affect the E-field [22] and hence activation thresholds [19], as the E-field can vary over distances comparable or larger than neuronal length constants. The mesoscopic E-field variations led to changes in activation thresholds of up to 30% in specific neuron models [9]. In contrast, microscopic E-field variations are spatially filtered by the neural membranes and average out on the mesoscale to mean values [10] that depend on the underlying cellular density and extracellular volume fraction [9]. Accounting for the microscopic structure [4], a model captured variations in the gray matter effective conductivity not only across layers but potentially also over short distances along the cortex within the same layer [9]. However, the assumption of such strong within- and across-layer conductivity variations needs to be experimentally validated, due to the modeling method simplifying the extracellular space as discussed below. Given the importance of mesoscale conductivity variations for E-field and neural simulations, we outline approaches bridging micro- and macroscale data to derive mesoscale conductivity distributions. Here we focus on low-frequency, quasistatic conductivities relevant for transcranial electrical and magnetic stimulation [23]; frequency-dependent and non-ohmic aspects of tissue impedance are important but largely separate from the mesoscale spatial questions highlighted here.

### *2.2. Macroscale approaches*

Higher resolution imaging can inform tissue models that capture mesoscale features, including meninges, large blood vessels, cortical layers, nuclei, or other variations in tissue properties within brain regions, and even microscopic structures within a region of interest [24]. However, electrical properties are mostly measured on macroscales and distributions on the mesoscale are not available for most tissue types.

Conventional metal microelectrodes and electrode arrays for impedance measurements average conductivity over relatively large volumes [16], whereas fine-tipped (~1 μm) glass micropipettes can resolve fine conductivity changes (Figs. 1 and 6B). Unlike the sharp tissue conductivity boundaries represented in models, actual transitions of tissue composition (e.g., from gray matter to white matter) are smoother [18,25]. Therefore, measurements must distinguish true conductivity gradients from artifacts introduced by the electrodes and tissue preparation (Fig. 1). Applying macroscale electric impedance tomography principles to multi-electrode arrays [26,27] could measure mesoscale resistivity distributions within brain tissue samples *in vitro*. Such systems require dense electrode coverage and careful calibration. Alternatively, multi-contact electrodes [14,22] can measure spatial E-field profiles during local microstimulation [28] or transcranial stimulation [22], informing conclusions about mesoscale conductivities and their spatial variations.

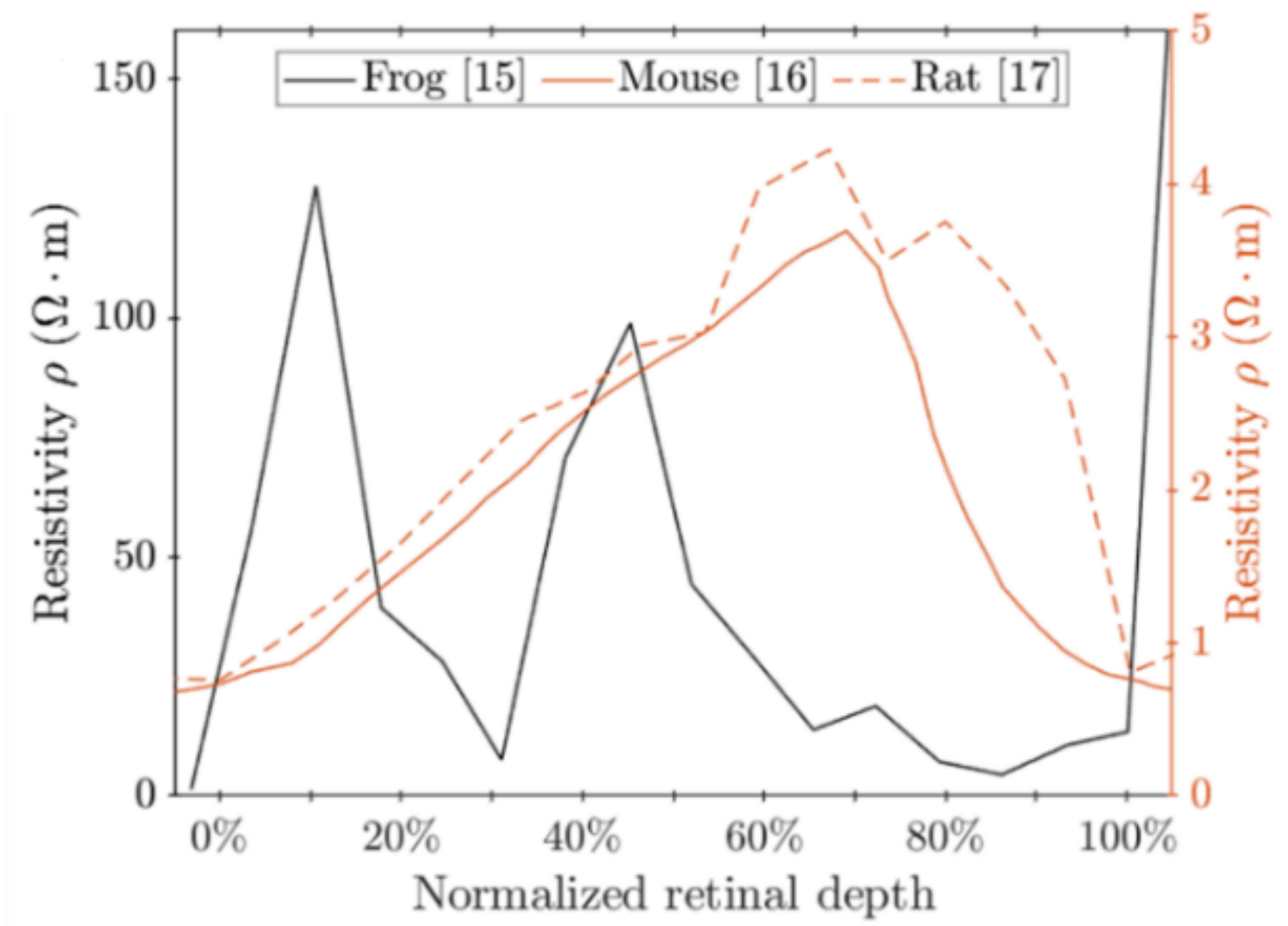

**Figure 1** Depth-dependent resistivity of the retina, with normalized depth of 0% indicating the retinal ganglion cell side and 100% indicating the photoreceptor side. Measurements using ion-selective micropipettes with sub-micrometer diameter in a frog eyecup (left axis, black line) show distinct resistivity variation with depth and much lower values due to the intact retinal pigment epithelium [15]. Measurements using microelectrodes in isolated rodent retina without the highly-resistive retinal pigment epithelium show lower resistivity and smooth profiles that average the conductivity of distinct layers and surrounding saline due to spatial averaging of the electrodes and tissue distortion and disruption during electrode insertion [16] (right axis, orange solid line: bipolar microelectrode with a conic inner pole of 25 μm diameter and 25 μm height and concentric outer pole [16]; orange dashed line: microelectrode array with 40 μm × 25 μm rectangular electrodes spaced 10 μm apart [17]).

### *2.3. Microscale approaches*

Microscale variations in tissue conductivity can be modeled using simplified geometries [31,32] and extracellular volume fraction data [30] or realistic tissue structure from electron microscopy [4,9,19]. However, most electron microscopy methods, including those used in recent large-scale connectomics projects [25,33], rely on chemical fixation, which causes tissue shrinkage and distortions of cellular structure due to dehydration [34–36]. Chemical fixation results in an almost complete absence of the 40–80 nm wide extracellular gaps between cell membranes [37], leading to direct contact between neighboring membranes [34] and underestimation of the extracellular volume fraction (Fig. 2A). This issue is mitigated in tissues processed with cryofixation, but to what extent the reconstructed cells are free from distortion is unknown. In a recent microscopic E-field model [6,7], the tissue had a volume fraction of "extracellular space" of around 70%, much larger than the 15%–30% range obtained from *in vivo* diffusion studies [30]. This discrepancy appears to arise because the model's "extracellular space" subsumed intracellular regions of omitted cell structures (e.g., neurites belonging to somata outside the sample as well as neuroglia; Fig. 2B). Thus, although these microscopic models provide valuable insight into E-field distortions around complex cellular (or vascular) morphologies, the uncertainties of extracellular space limit the use of these models for conductivity estimation. Super-resolution optical microscopy enables nanoscale reconstruction of brain tissue in living samples while avoiding fixation- or freezing-induced artifacts [38], but these methods are limited to shallow depths due to light scattering and are still not viable *in vivo*. With enhancements of these techniques, structural imaging could be combined with physiological measurements in the same preparation to build microscale reconstructions that accurately capture the extracellular space and enable accurate determination of micro- and meso-scale tissue conductivities.

### *2.4. Modeling layer-specific cortical conductivity*

Laminar probe measurements during transcranial electrical stimulation revealed marked variation in E-field amplitude across cortical layers [22]. Since the current density is relatively uniform beneath the large electrodes, this indicates conductivities vary substantially between cortical layers. Therefore, obtaining layer-specific conductivity measurements in the cortex would improve modeling accuracy, especially for invasive stimulation using electrodes for which the local field gradients are high. To obtain mesoscopic conductivity values from microscopic cellular structures, we examine how the conductivity of neural tissue varies as a function of cell density and thus volume fraction using computational models. As layer-specific cellular volume fractions are not directly available, complementary volume fractions of extracellular space from diffusion experiments were used.

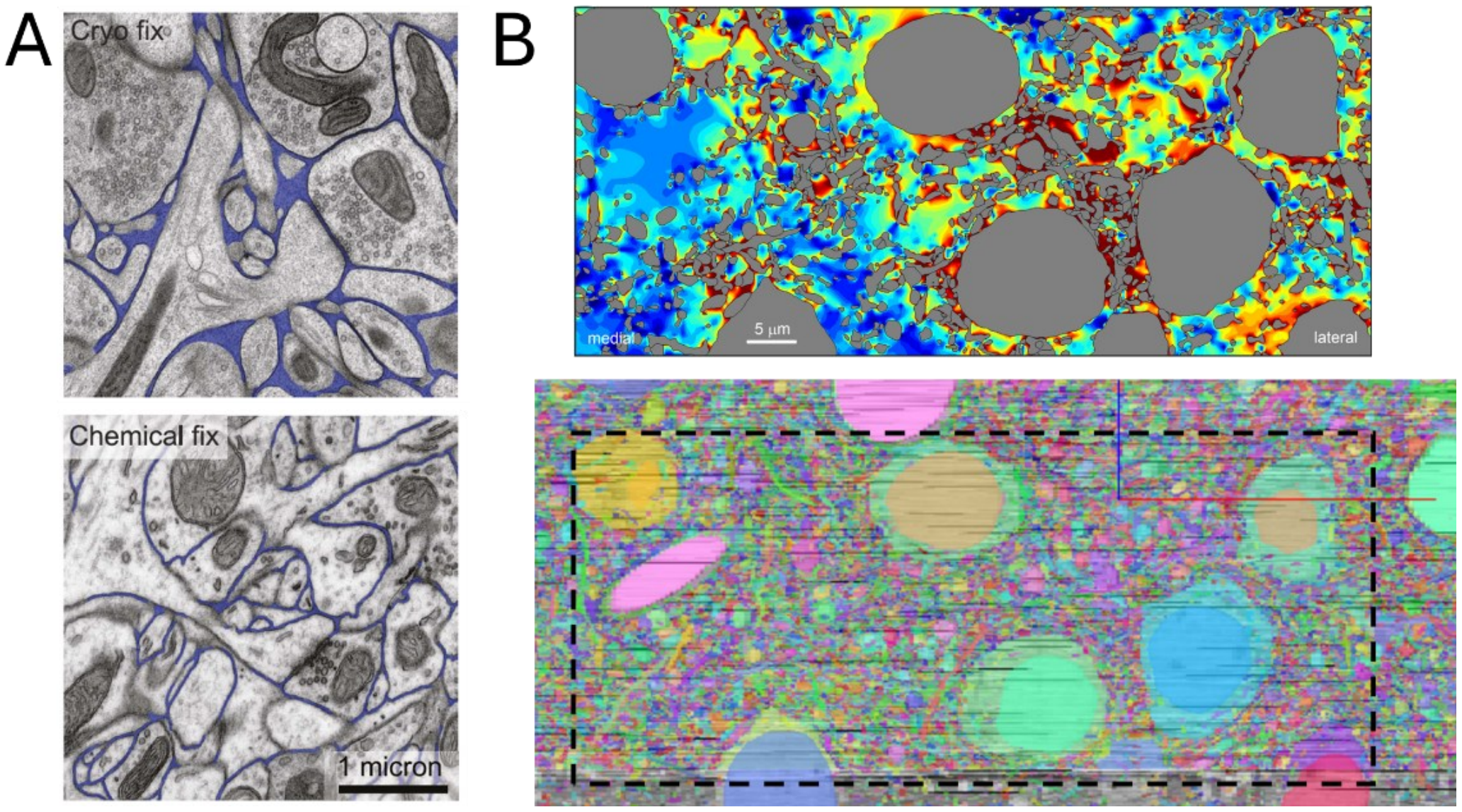

**Figure 2** Challenges for accurate representation of extracellular space in tissue preparation for microscopic models. **A.** Electron microscopy images of tissue processed with cryofixation (top) and chemical fixation (bottom) show significant difference in extracellular space (pseudo-colored in blue) among other fixation-induced changes [34]. **B.** Top: Microscopic E-field simulation [6,7] showing large extracellular space (color map showing E-field) with ~70% volume fraction and small intracellular space (gray) of ~30% volume fraction. Bottom: Electron microscopy of the MICrONS dataset of mouse visual cortex [33] on which the microscopic E-field simulations were based showing the same frame (dashed outline). The micrograph reveals dense cellular and subcellular composition (filled with random colors) with almost no extracellular space due to chemical fixation, indicating that the "extracellular space" in the simulation is dominated by the intracellular space of excluded cells. Panel A reprinted from Ultrastructural Analysis of Adult Mouse Neocortex Comparing Aldehyde Perfusion with Cryo Fixation, by N. Korogod et al., 2015, eLife, vol. 4, p. e05793. Copyright 2015 by The Authors. Adapted under the Creative Commons Attribution License (CC BY 4.0). Top part of panel B adapted from Statistical Method Accounts for Microscopic Electric Field Distortions Around Neurons When Simulating Activation Thresholds, by K. Weise et al., 2025, Brain Stimulation, vol. 18, no.2, p. 281. Copyright 2025 by The Authors. Adapted under the Creative Commons Attribution License (CC BY-NC 4.0). Bottom part of panel B obtained from www.microns-explorer.org.

## 3. Methods

### *3.1. Theoretical formulas for conductivity–cell volume fraction relationship*

Previous studies [39–42] modeled the relationship between effective conductivity $\sigma$ and volume fraction of cells $f$ (Table 1). Most methods assumed a low volume fraction for cell suspensions, with the conductivity of the extracellular fluid as $\sigma_0$ and homogenous cell conductivity, which for simplicity can be set to zero given the extremely low conductivity of cell membranes.

**Table 1** Theoretical conductivity–cell volume fraction relationship

| Formula | Notes | Sources |
|---|---|---|
| $\frac{\sigma}{\sigma_0}=\frac{1-f}{1+f/2}=1-\frac{3f}{2+f}$ | Dilute cell suspension, simplified for non-conducting spheres | [39,42] |
| $\frac{\sigma}{\sigma_0}=1-\frac{3f}{2+f-0.394\cdot f^{\frac{10}{3}}}$ | Spherical particles arranged in a simple cubic lattice, simplified for non-conducting spheres. | [40,42] |
| $\frac{\sigma}{\sigma_0}=1-\frac{3f}{2+f-\frac{0.986\cdot f^{\frac{10}{3}}}{1-0.307\cdot f^{\frac{7}{3}}}}$ | Spherical particles arranged in a simple cubic lattice, simplified for non-conducting spheres. | [43] |
| $(1-f)^{\frac{5}{3}}\leq\frac{\sigma}{\sigma_0}\leq(1-f)^{\frac{3}{2}}$ | Elongated homogeneously distributed randomly orientated non-conducting spheroids | [41] |
| $\frac{\sigma}{\sigma_0}=\begin{cases}(1-f)^1\\(1-f)^2\end{cases}$ | Parallel fibers, longitudinal<br>Parallel fibers, transverse | [41] |

### *3.2. Layer-specific volume fraction of extracellular space and cells in brain*

We used the volume fraction of extracellular space, $\alpha$, obtained by Lehmenkühler et al. with *in vivo* diffusion methods in somatosensory cortex and subcortical white matter of adult rats (Table 2) [30]. In summary, tetramethylammonium concentration was measured with ion-selective microelectrodes positioned 130–200 µm from an iontophoretic source and diffusion equations were fit to the concentration versus time assuming spherical symmetry to extract parameters including $\alpha$, tortuosity, and cellular uptake. The volume fractions of cells were given as $f=1-\alpha$.

### *3.3. Tissue models and simulation setup*

We built cubic blocks of simplified microscopic cellular representations of several tissue types using finite element models in COMSOL (version 6.2, COMSOL, Inc.). Three tissue types were modeled: parallel axons (2D), cell bodies (3D), and interleaved neurites (3D). Conductivities of cell membranes and the intra- and extra-cellular spaces were assigned according to literature (Table 3). The cell volume fraction varied in steps of 5% from 5% to 95% or within the maximum allowed by the specific tissue model (Table 4), and different cell shapes and orientations of the tissue structure were modeled to quantify their influence on the conductivity. The potential and E-field distributions were simulated for 1 µA current applied on one side of the sample with a floating potential boundary condition. The opposite side was grounded, and the remaining four sides were insulated. The resulting electrode floating potential was used to calculate the total resistance of the tissue block. The effective tissue conductivity for a given tissue type was then calculated and averaged for various cell shapes and rotational orientations of the tissue structure. The conductivities of different cortical layers (layers 2 to 6) were interpolated based on experimental volume fraction in the range of 70% to 90%. Data analysis and additional visualization were conducted in MATLAB (version 2024b, The MathWorks, Inc.).

**Table 2** Volume fraction of extracellular space and cells [30]

| Layer | Volume fraction of extracellular space, $\alpha$ | Volume fraction of cells, $f$ |
|---|---|---|
| L1 | Not available | Not available |
| L2 | 17% | 83% |
| L3 | 20% | 80% |
| L4 | 21% | 79% |
| L5 | 22% | 78% |
| L6 | 24% | 76% |
| White matter | 18% | 82% |

*3.3.1. Parallel axons*

Cylinders and hexagonal prisms [44,45] placed in hexagonal grids were used to represent parallel axons (Fig. 3A and Table 4). The length of the unit cell (parallelogram) was 10 μm. Circular cross-section allows cell volume fractions up to 90.7%, whereas hexagonal prisms allow denser packing theoretically achieving 100% volume fraction. Given the longitudinal symmetry along the axons' axes, a 2D model was built with the axial direction not explicitly represented. The tissue blocks were 180 × 180 × 180 $\mu m^3$ in size, and the orientation of the grid pattern was rotated up to 30° in steps of 5°.

*3.3.2. Cell bodies*

Spheres [42] and truncated octahedrons [46] placed in body-centered cubic lattice with a unit cell of 10 μm side length were used to represent tight packing of cells (Fig. 3B and Table 4), with the former being limited to a volume fraction of 68%. The 3D tissue block was 60 × 60 × 60 $\mu m^3$ in size and three tissue orientations, corresponding to Miller indices of 100, 110, and 111 of the lattices, were used.

*3.3.3. Interleaved neurites (neuropil)*

3D-crosses with circular and square cross-sections placed in body-centered cubic lattice were used to represent densely interleaved neural branches in the neuropil (Fig. 3C and Table 4), with the 3D model setup similar to that of the cell bodies.

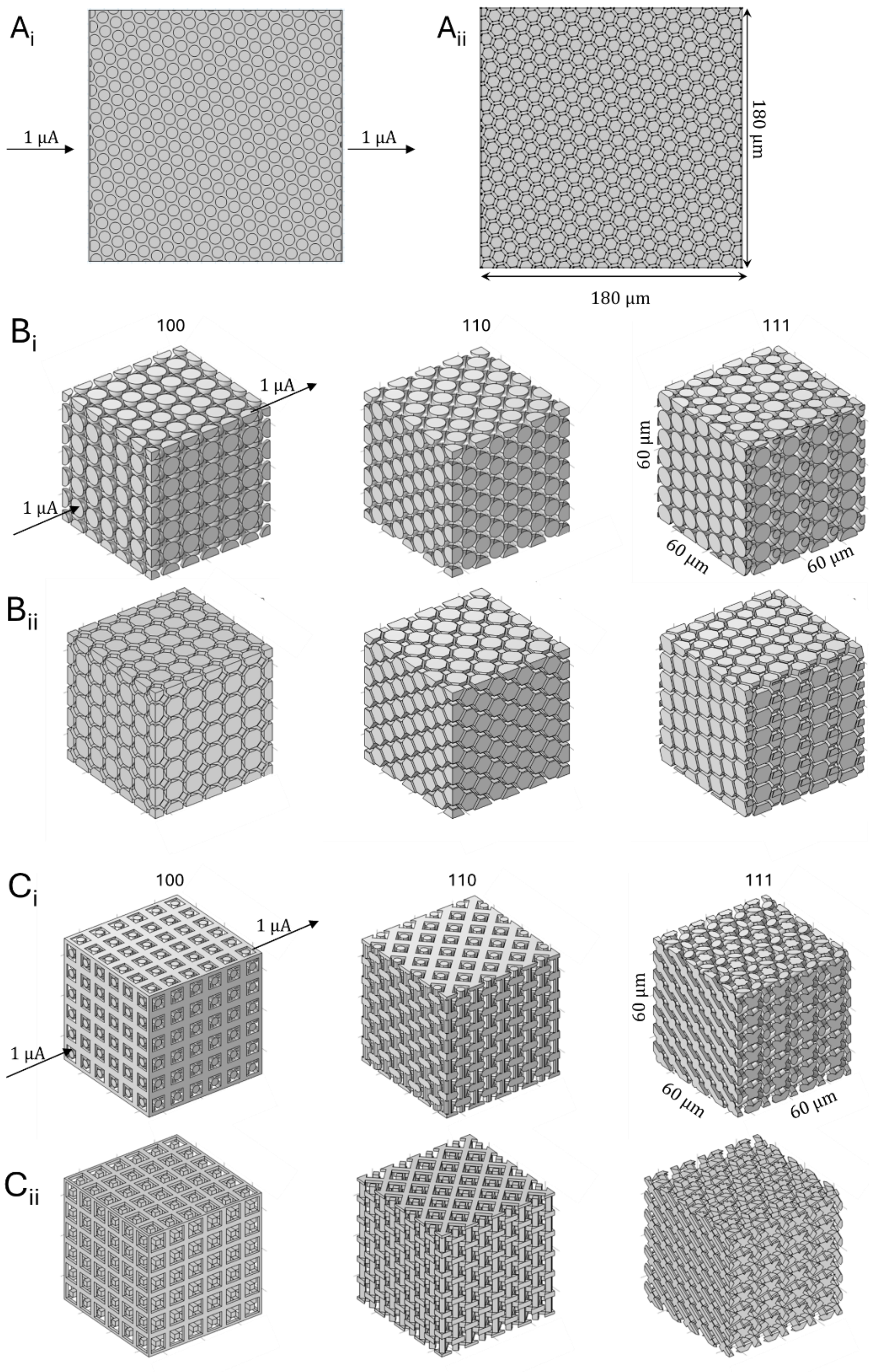

**Figure 3 A.** Axons transverse to the current direction modeled as hexagonal grids of cylinders (**A**$_{\mathbf{i}}$) and hexagon prisms (**A**$_{\mathbf{ii}}$). The 2D model had an implicit depth of 180 μm. Both axon grids had a cell volume fraction of 60% and rotation of 20°. 0.5 μm-wide gaps were added on both the left and right sides to avoid direct contact between the electrodes and cell membrane. **B.** Cells modeled as body-centered cubic lattices of spheres (**B**$_{\mathbf{i}}$) and truncated octahedrons (**B**$_{\mathbf{ii}}$). The current was aligned with the 100 (left), 110 (middle), and 111 (right) orientations of lattices, all with a cell volume fraction of 60%. The outer boundaries of the models are not shown. 0.5 μm-wide gaps were added on both the left and right sides to avoid direct contact between the electrodes and cell membrane. **C.** Interleaved neurites modeled as body-centered cubic lattices of 3D-crosses of circular (**C**$_{\mathbf{i}}$) and square (**C**$_{\mathbf{ii}}$) cross-section, with cell volume fractions of 48.5% and 31.3%, respectively. Other modeling setups are similar as for the cell models in Panel B.

**Table 3** Model parameters

| Parameter | Symbol | Value | Unit |
|---|---|---|---|
| Extra- and intracellular conductivity | $\sigma_0$ | 1 | S·m$^{-1}$ |
| Membrane conductivity | $\sigma_\mathrm{m}$ | $10^{-8}$ | S·m$^{-1}$ |
| Membrane thickness | $t_\mathrm{m}$ | 10 | nm |
| Specific membrane resistance | $R_\mathrm{m}$ | 1 | Ω·m$^2$ |
| Unit cell side length | $d$ | 10 | μm |

**Table 4** Models and parameters for representing different neural tissue types

| Tissue type (model dimension) | Lattice with unit cell of side length $d$ and orientations explored | Shape of cells or their cross sections, Side length $\hat{a}$ or radius $\hat{r}$ (normalized to $d$) | | Volume fraction $f$ and limits (if less than 100%) | Inverse relationship of volume fraction and neuron size |
|---|---|---|---|---|---|
| Parallel fibers (2D, transverse) | Hexagon grid, 0° to 30°, in steps of 5° | Hexagon | $\hat{a} \leq \frac{1}{\sqrt{3}}$ | $f = 3\hat{a}^2$ | $\hat{a} = \sqrt{\frac{f}{3}}$ |
| | | Circle | $\hat{r} \leq \frac{1}{2}$ | $f = \frac{2\pi}{\sqrt{3}}\hat{r}^2 \leq 90.7\%$ | $\hat{r} = \sqrt[4]{3}\sqrt{\frac{f}{2\pi}}$ |
| Cell bodies (3D) | Body-centered cubic lattice, Miller-indices 100, 110, 111 | Truncated octahedron | $\hat{a} \leq \frac{1}{2\sqrt{2}}$ | $f = \left(2\sqrt{2}\hat{a}\right)^3$ | $\hat{a} = \frac{\sqrt{2}\sqrt[3]{f}}{4}$ |
| | | Sphere | $\hat{r} \leq \frac{\sqrt{3}}{4}$ | $f = \frac{8\pi}{3}\hat{r}^3 \leq 68.0\%$ | $\hat{r} = \sqrt[3]{\frac{3f}{8\pi}}$ |
| Dense interleaved neurites (3D) | | 3D cross with length $d$ and square cross section | $\hat{a} \leq \frac{1}{2}$ | $f = 2(3 - 2\hat{a}) \cdot \hat{a}^2$ | $\hat{a} = \cos\left[\frac{\mathrm{acos}(1-f) - 2\pi}{3}\right] + \frac{1}{2}$ |
| | | 3D cross with length $d$ and circular cross section | $\hat{r} \leq \frac{1}{4}$ | $f = \left(6\pi - 16\sqrt{2}\hat{r}\right) \cdot \hat{r}^2 \leq 82.5\%$ | $\hat{r} = \frac{\pi}{4\sqrt{2}}\left\{\cos\left[\frac{1}{3}\mathrm{acos}\left(1 - \frac{32f}{\pi^3}\right) - \frac{2\pi}{3}\right] + \frac{1}{2}\right\}$ |

## 4. Results

### *4.1. Field distributions*

The potential distribution varied linearly from one electrode to the other, whereas the E-field was very high in the extracellular space and extremely low inside the cells due to the highly resistive membrane (Fig. 4, showing representative examples for 2D and 3D models).

### *4.2. Effective conductivity*

The effective conductivities calculated from the total resistance of the tissue blocks for a given cell volume fraction were averaged across different orientations of the tissue structure relative to the main direction of the E-field (Fig. 5). The coefficient of variation for each combination of tissue type and cell volume fraction was less than 0.03, showing that the orientation of the tissue structure had no impact on the effective conductivity. Further, the conductivity of different cell shapes for the same tissue type agreed with each other, indicating that the effective conductivity–volume-fraction relationships were not affected substantially by cell shape either.

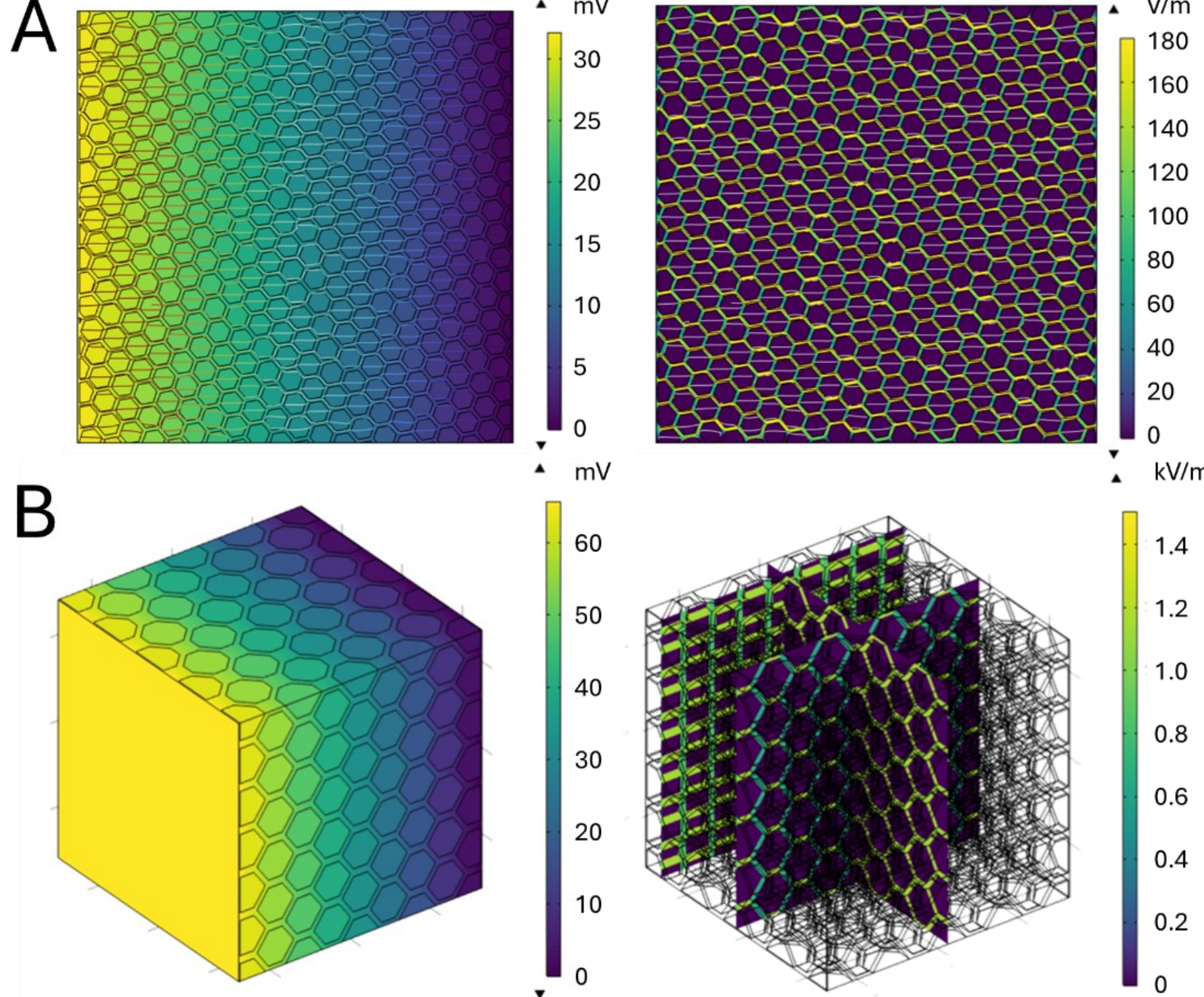

**Figure 4** Example potential (left) and E-field amplitude (right) distributions. **A.** 2D model of hexagonal axon grid with 70% cell volume fraction. **B.** 3D model of truncated octahedron cells with 65% cell volume fraction.

The effective tissue conductivities for the three modeled tissue structures were monotonically decreasing convex functions of the cell volume fraction. They followed the theoretical power formula [41] only for small volume fractions (< 20%) and deviated to higher values for larger volume fractions. The convexity of the relationship reduces differences in cell volume fraction for high values, resulting in smaller conductivity contrasts than assumed from a linear relationship [9]. Surprisingly, Maxwell's formula, which was originally developed for dilute cell suspensions, matched the relationship for cells bodies in the entire range of volume fractions, and an empirical relationship inspired by Maxwell's formula fitted the axonal transverse conductivity (Fig. 5).

### *4.3. Layer-specific conductivity*

In the cortex and subcortical white matter, approximately 75% of the cellular volume consists of elongated structures such as axons, dendrites, cilia, and blood vessels (see Table S4 of [25]). Therefore, we interpolated the conductivity of each cortical layer from the simulated results for interleaved neurites (Fig. 5B) using the layer-specific cell volume fractions (Table 2), with the results shown in Fig. 6A. Here, the conductivity of the extracellular space was assumed to be cerebrospinal fluid, 1.8 S/m [47]. The conductivity of subcortical white matter was also estimated for isotropic and anisotropic cases, with the latter using the simulated results for transverse conductivity of parallel axons (Fig. 5A) and empirical formula for longitudinal conductivity (Table 1).

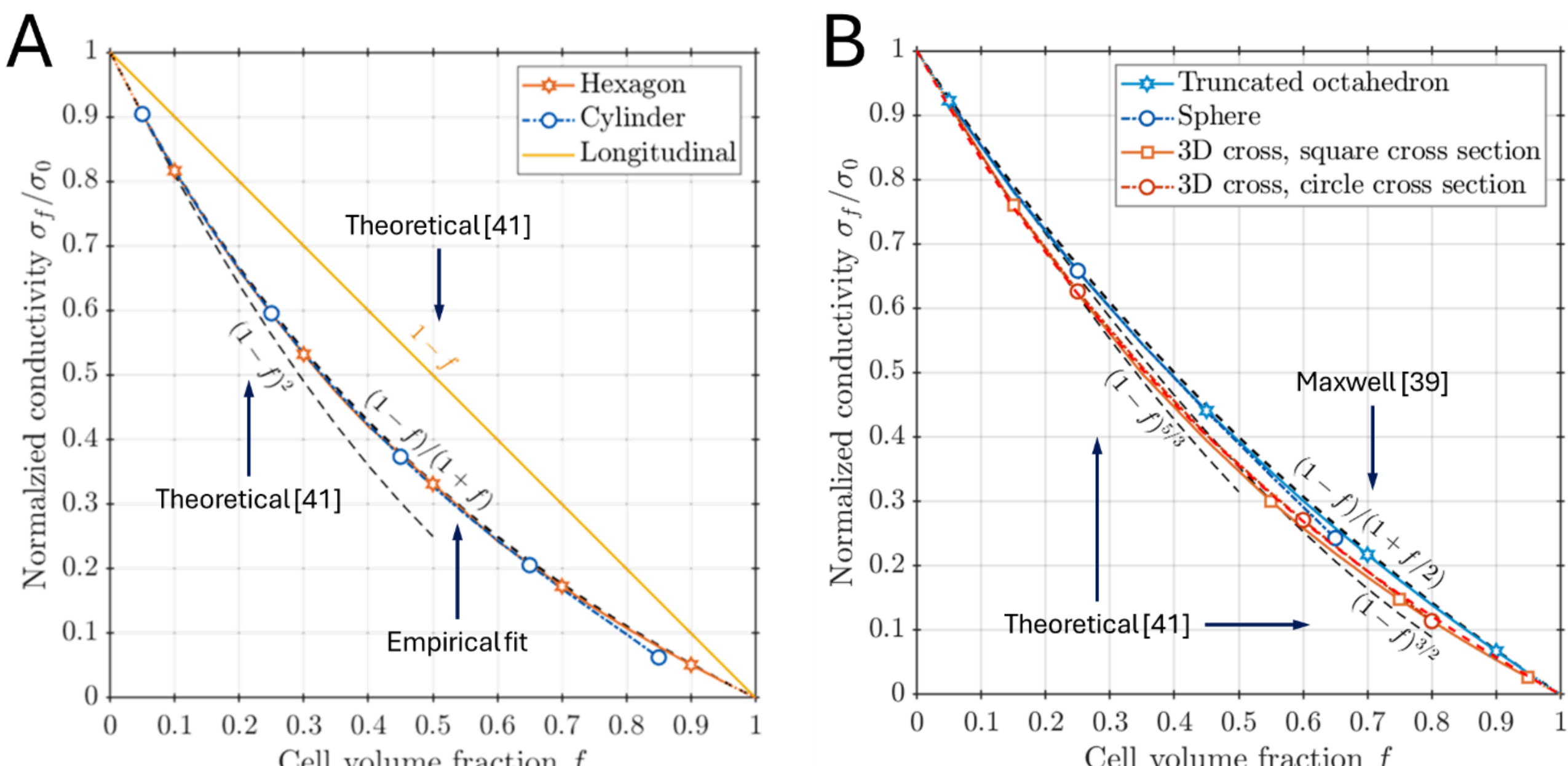

**Figure 5** Effective tissue conductivity normalized to extracellular fluid as function of volume fraction and averaged across orientations. **A.** 2D models (average of 7 orientations). **B.** 3D models (average of 3 orientations). Conductivity is not affected by the cell shape as demonstrated by the overlapping lines for the same tissue type. Theoretical formulas and empirical fits are also plotted as dashed lines.

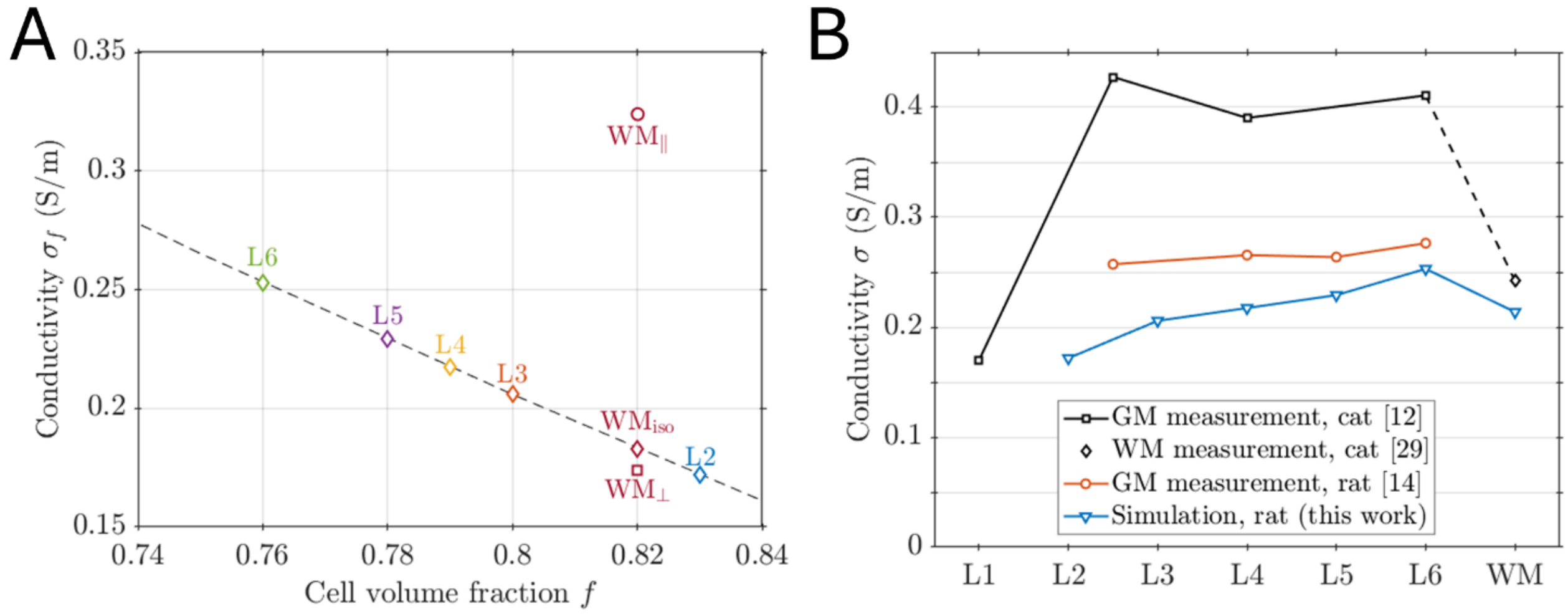

**Figure 6 A.** Effective tissue conductivity of cortical layers shown according to their cell volume fraction, assuming $\sigma_0$ as 1.8 S/m. Markers are interpolated according to experimental data from [30]. The dashed line corresponds to simulated results for interleaved neurites (Fig. 5). Subcortical white matter conductivity was also calculated for isotropic and anisotropic cases based on its volume fraction. **B.** Equivalent isotropic conductivity of gray matter measured in the cat somatosensory cortex (black squares) using glass micropipettes [12], with layers 2 and 3 considered as a single layer and layer 5 data not collected, and equivalent isotropic conductivity of white matter measured in the cat internal capsule (black diamond) using a wire probe [29]. It should be noted that for layer 1, only the vertical conductivity across layer was available, whereas other layers also had conductivities in both the mediolateral and anterior–posterior directions. Equivalent isotropic conductivity of gray matter measured in the rat somatosensory cortex (orange circles) using electrode arrays with 15 and 30 µm diameter electrodes [14], with layers 2 and 3 considered as a single layer. The isotropic conductivity estimated using computational modeling (blue triangles, interpolated data in panel A) based on extracellular volume fraction in rat somatosensory cortex and subcortical white matter obtained from ion diffusion measurements with ion-sensitive glass microelectrodes of ~1-µm diameter [30].

With decreasing cell volume fraction, the conductivity of cortical layers increased with depth from layer 2 to layer 6 (Fig. 6A). Conductivity of layer 1 was estimated to be the lowest (0.14 S/m [18]), in agreement with this trend. Although the variation of conductivity within the cortex was small when compared to the conductivity of the cerebrospinal fluid (9% to 15%), the conductivity difference was considerably larger when compared between layers, e.g., with layers 3 and 6 respectively 20% and 50% more conductive than layer 2.

With high cell volume fractions, the simulated isotropic conductivity of subcortical white matter was lower than all layers except layer 2, although it should be noted there were substantial differences in the longitudinal and transverse conductivities as expected with anisotropy, with longitudinal conductivity measured as high as 1.1 S/m in some studies [29].

## 5. Discussion

In this work we first conducted theoretical analyses of the spatial scales of tissue properties and E-field in brain stimulation, highlighting the importance of the mesoscale in accurate modeling. Approaches that estimate effective mesoscopic tissue conductivities from micro- and macroscales complement and validate each other and should converge on consistent mesoscale distributions. The technical challenges of these approaches include limited spatial resolution of the conductivity due to the size and number of the measurement electrodes and tissue deformation and shrinkage during chemical fixation for electron microscopy. These challenges need to be addressed via collaboration across different fields, including biophysics and physiology, electrode microfabrication, high resolution imaging, histology, and computational modeling. Further combining mesoscale models with physiological experiments could identify sufficient levels of biophysical detail for predictive models and validate statistical averaging across neurons within a tissue volume and across subjects.

Demonstrating the microscale approach, we built microscopic tissue models with simplified cell geometry to evaluate effective tissue conductivity using cell volume fraction data obtained from diffusion measurements. Compared to realistic cellular models obtained with electron microscopy, the simple models allowed precise control of the volume fraction of the cells and extracellular volume homogenized on the mesoscale, and circumvented issues of realistic models including chemical-fixation-induced tissue shrinkage and high computational costs. For a given tissue type, the effective conductivity–volume-fraction relationships of the spatially symmetric models were not affected by orientation of the E-field relative to the tissue structure or the cell shapes we tested. Whether such cell shape independence holds for realistic cell morphologies should be further examined with more complex models [25,33]. Except for Maxwell's formula, prior theoretical results did not agree with the relationship at high volume fractions relevant for brain tissue.

We estimated layer-specific cortical and subcortical conductivity using an interpolation of the simulated effective conductivity for interleaved neurites and assuming the extracellular conductivity to equal that of the cerebrospinal fluid. The relative difference in conductivity between the layers was substantial. The results agreed with the higher E-field measured in layer 2/3 compared to those in layers 4 and 5/6 during transcranial alternating current stimulation (tACS) [22], in which the large surface electrode resulted in relatively uniform current density in the cortex, and thus the E-field was inversely proportional to conductivity. The modeled conductivity also exhibited a similar trend across layers compared to the equivalent isotropic conductivity of gray matter measured in the rat somatosensory cortex (Fig. 6B, orange circles [14]), albeit with smaller absolute values. Such differences may be attributed to an underestimation of the extracellular/interstitial fluid conductivity by using that of cerebrospinal fluid for the effective extracellular conductive space.

The tissue models had several limitations besides simplified cell geometry. Realistic tissue consists of a mixture of cell sizes and types, with irregular shapes and higher tortuosity for the current flow due to dead ends and concave size that require more complex cell shapes to represent. The accuracy of the modeled cortical conductivity also depended on that of the volume fraction of extracellular space. The experimental volume fractions used in this study were obtained by fitting the measured ion concentration to isotropic diffusion equations, which did not account for anisotropy [30]. Therefore, anisotropic conductivity values could be obtained with improved experimental estimation of volume fractions and further modification of the models to use asymmetric geometries (e.g., spheroids [41] instead of spheres). Cell membrane capacitance can also be modeled to arrive at frequency-dependent conductivity and capacitance. Ultimately, realistic cell structure from electron microscopy without tissue deformation should be utilized to arrive at mesoscale conductivity distribution of different cortical layers and brain regions.

## 6. Conclusions

Despite the numerous challenges of both microscale and macroscale approaches, they should converge on consistent mesoscale distributions of tissue conductivity. We demonstrated a simple mathematical method to evaluate tissue conductivity using cell volume fraction data, which can be obtained from microscopic tissue samples. Overcoming technical and methodological challenges, the quality of E-field models and simulations of brain stimulation can be improved by accurate representation of tissue conductivity on the mesoscale, including within the gray matter and at the transition from gray matter to white matter. Tissue conductivities vary substantially between cortical layers, and using layer-specific values within the cortex could improve the accuracy of E-field models for brain stimulation.

**Acknowledgments**

This work was supported by the National Institute of Mental Health and the National Institute of Neurological Disorders and Stroke of the US National Institutes of Health under grant No. R01MH128422 and U01NS126052. A. Thielscher was supported by the Lundbeck Foundation grant R313-2019-622. The content is solely the responsibility of the authors and does not necessarily represent the official views of the funding agency. Preliminary results were presented at the Neural Interfaces 2025 conference, Arlington, VA, USA, June 12–14, 2025 [48].

**CRediT authorship contribution statement**

B. Wang: Conceptualization, Methodology, Investigation, Formal analysis, Visualization, Writing–original draft, Writing–review & editing. T. H. Worbs: Investigation, Formal analysis, Visualization, Writing–review & editing. M. A. Hussain: Investigation, Writing–review & editing. A. S. Aberra: Investigation, Writing–review & editing. A. Thielscher: Conceptualization, Funding acquisition, Writing–review & editing. W. M. Grill: Conceptualization, Funding acquisition, Writing–review & editing. A. V. Peterchev: Conceptualization, Funding acquisition, Supervision, Project administration, Writing–review & editing.

**Conflict of interest**

B. Wang, T. H. Worbs, M. A. Hussain, A. S. Aberra, A. Thielscher, and W. M. Grill declare no relevant conflict of interest. A. V. Peterchev has received patent royalties and consulting fees from Rogue Research; equity options, scientific advisory board membership, and consulting fees from Ampa Health; equity options, consulting fees, and travel support from Magnetic Tides; consulting fees from Soterix Medical; equipment loans from MagVenture; hardware donations from Magstim; and research funding from Motif.